\documentclass[aip,preprint]{revtex4-2}
\usepackage[cp1251]{inputenc}
\usepackage[T2A]{fontenc}
\usepackage[english]{babel}
\usepackage{amssymb,latexsym,amsmath,amscd}
\usepackage{graphicx,color,framed}
\usepackage{xcolor}
\usepackage{siunitx}
\usepackage{empheq}
\usepackage{xr}
\usepackage{microtype}
\usepackage{amsfonts}
\usepackage[normalem]{ulem}
\graphicspath{{figures/}}

\makeatletter
\newcommand*{\addFileDependency}[1]{
  \typeout{(#1)}
  \@addtofilelist{#1}
  \IfFileExists{#1}{}{\typeout{No file #1.}}
}
\makeatother
\newcommand*{\myexternaldocument}[1]{%
    \externaldocument{#1}%
    \addFileDependency{#1.tex}%
    \addFileDependency{#1.aux}%
}
\myexternaldocument{./supporting_information}

\begin{document}
\title{Macroscopic forces in inhomogeneous polyelectrolyte solutions}
\author{\firstname{Yury A.} \surname{Budkov}}
\email[]{ybudkov@hse.ru}
\affiliation{School of Applied Mathematics, HSE University, Tallinskaya st. 34, 123458 Moscow, Russia}
\affiliation{G.A. Krestov Institute of Solution Chemistry of the Russian Academy of Sciences, 153045, Akademicheskaya st. 1, Ivanovo, Russia}
\author{\firstname{Nikolai N.} \surname{Kalikin}}
\affiliation{School of Applied Mathematics, HSE University, Tallinskaya st. 34, 123458 Moscow, Russia}
\affiliation{G.A. Krestov Institute of Solution Chemistry of the Russian Academy of Sciences, 153045, Akademicheskaya st. 1, Ivanovo, Russia}

\begin{abstract}
In this paper, we present a self-consistent field theory of macroscopic forces in spatially inhomogeneous flexible chain polyelectrolyte solutions. We derive an analytical expression for a stress tensor which consists of three terms: isotropic hydrostatic stress, electrostatic (Maxwell) stress, and stress rising from conformational entropy of polymer chains -- conformational stress. We apply our theory to the description of polyelectrolyte solutions confined in a conductive slit nanopore and observe anomalous behavior of disjoining pressure and electric differential capacitance.
\end{abstract}
\maketitle
\section{Introduction}
Modern electrochemical devices, such as batteries and supercapacitors, extensively utilize porous electrodes impregnated with low molecular weight electrolyte solutions or room temperature ionic liquids (RTILs) \cite{zhang2021review,liu2019three}, although from general considerations one can expect a higher electric double layer charge when dealing with long polyelectrolyte chains. We have recently proposed \cite{budkov2022electrochemistry,kalikin2022polymerized} a theoretical model describing charged polymer chains near an electrified electrode and observed a substantial increase in differential capacitance values when considering the case of their solution in a polar organic solvent.

Turning back to the case of low molecular weight charge carriers, electrosorption of small ions into porous materials is known to be accompanied by deformation of the latter~\cite{kolesnikov2022electrosorption}, which in turn is closely related to such an important quantity as disjoining pressure if we talk about slit pores or solvation pressure -- for pores of an arbitrary geometry~\cite{kolesnikov2021models,gor2017adsorption}. Thus, an obvious extension of the above-mentioned work is an investigation of both the disjoining pressure and the differential capacitance of a polyelectrolyte solution in a slit pore with conductive walls in order to get new insights into supercapacitors.

However, there have hardly been any works devoted to studying the disjoining pressure in nanopores, even for conventional electrolyte solutions and RTILs~\cite{de2022structural,misra2019theory,budkov2022modified,kolesnikov2022electrosorption}. To the best of our knowledge, there have been no studies investigating the disjoining pressure of polyelectrolyte solutions confined in conductive nanopores as a function of pore size and surface electrostatic potential either. Nevertheless, it is worth noting a number of important theoretical works dealing with adsorption of both single polyelectrolyte chains~\cite{muthukumar1987adsorption,podgornik2004polyelectrolyte,cherstvy2012polyelectrolyte,brilliantov2016generation} and multiple chains from a solution~\cite{netz2003neutral,joanny1999polyelectrolyte,landman2021repulsive} onto oppositely charged surfaces of membranes or colloidal particles. Despite that, none of these papers proposed a systematic approach to the calculation of macroscopic forces acting on a dielectric or conducting body immersed into a polyelectrolyte solution or melt.

Below we propose a self-consistent field theory of macroscopic forces in inhomogeneous flexible chain polyelectrolyte solutions. As a special case, we apply it to the investigation of the disjoining pressure in a polyelectrolyte solution confined in a slit conductive nanopore. We also show how disjoining pressure behavior manifests itself on differential capacitance profiles.

\section{Theory}
Let us consider a polyelectrolyte solution consisting of polymerized flexible cations (macroions), whose monomeric units carry a charge $q>0$ and low-molecular-weight anions (counterions) with a charge $-q$. Note that promising polyelectrolyte materials such as polymeric ionic liquids have positively charged macroions and negatively charged counterions~\cite{budkov2022electrochemistry,kalikin2022polymerized}. We assume the polymerization degree of macroions to be very high ($N\gg 1$), which means we can neglect the translation entropy of the mass center of the polymer chains. The grand thermodynamic potential (GTP) of the solution is
\begin{equation}
\label{GTP}
\Omega=\int d\bold{r}\omega(\bold{r}),
\end{equation}
where we have introduced the GTP density
\begin{equation}
\label{omega_2}
\omega=-\frac{\varepsilon(\nabla\psi)^2}{2}+\rho\psi +\frac{k_{B}Tb^2}{6}(\nabla n_{p}^{1/2})^2+f-\mu_{p}n_{p}-\mu_{c}n_c.
\end{equation}
The first and second terms in the integrand are the electrostatic energy density in the mean-field approximation with the local charge density $\rho(\bold{r})=q\left(n_{p}(\bold{r})-n_{c}(\bold{r})\right)$
and electrostatic potential $\psi(\bold{r})$; $\varepsilon$ is the dielectric permittivity of the solvent; the third term is the density of the conformational free energy \cite{lifshitz1969some,khokhlov1994statistical} of the flexible polymer chains with a bond length $b$ ($k_{B}$ is the Boltzmann constant, $T$ is the temperature). The fourth term, $f=f(n_{p},n_{c})$, determines the contribution of the volume interactions of monomeric units and counterions to the total free energy density, with $n_{p,c}(\bold{r})$ being the local concentrations of monomeric units and counterions, which we describe within the lattice model (without attractive interactions)~\cite{borukhov1997steric,kornyshev2007double,Maggs2016}
$f=k_{B}Tv^{-1}\left(\phi_c\ln\phi_{c}+\left(1-\phi_{c}-\phi_{p}\right)\ln\left(1-\phi_{c}-\phi_{p}\right)\right)$, where $\phi_{p,c}=n_{p,c}v$ are the local volume fractions of the counterions and monomeric units, $v$ is the elementary cell volume that is related to the bond length via the natural condition, $v=b^3$; $\mu_{p}$ and $\mu_c$ are the bulk chemical potentials of the monomeric units and counterions, respectively. The self-consistent field equations, which are simply the Euler-Lagrange equations for functional (\ref{GTP}), are
\begin{equation}
\label{EL_eq}
\frac{\partial{\omega}}{\partial{\psi}}=\partial_{i}\frac{\partial \omega}{\partial(\partial_{i}\psi)},~\frac{\partial{\omega}}{\partial{n_{p}^{1/2}}}=\partial_{i}\frac{\partial \omega}{\partial(\partial_{i}n_{p}^{1/2})},~\frac{\partial{\omega}}{\partial{n_{c}}}=0,
\end{equation}
where $\partial_i=\partial/\partial{x}_{i}$ is the partial derivative with respect to the Cartesian coordinates $x_i$ ($i=1,2,3$). Note that we adopted the Einstein rule implying the summation over the repeated indices. Using GTP density (\ref{omega_2}) introduced above, we arrive at
\begin{empheq}[left=\empheqlbrace]{align}\nonumber
\label{scf_eq}
&\bar{\mu}_{c}(\bold{r})-q\psi(\bold{r})=\mu_{c}\\
&\bar{\mu}_{p}(\bold{r})-\frac{k_{B}Tb^2}{6n_{p}^{1/2}(\bold{r})}\nabla^2 n_{p}^{1/2}(\bold{r})+q\psi(\bold{r})=\mu_{p}\\\nonumber
&\nabla^2\psi(\bold{r})=-\frac{q}{\varepsilon}\left(n_{p}(\bold{r})-n_{c}(\bold{r})\right),
\end{empheq}
where 
$\bar{\mu}_{c}={\partial f}/{\partial n_{c}}=k_{B}T\ln\left({\phi_c}/{(1-\phi_c-\phi_p})\right)$,
$\bar{\mu}_{p}={\partial f}/{\partial n_{p}}=-k_{B}T\ln\left(1-\phi_{c}-\phi_{p}\right)$
are the intrinsic chemical potentials of monomeric units and counterions, respectively. Taking into account that in the bulk solution, where $\psi=0$, the local electroneutrality condition, $n_{p}=n_c=n_0$, is fulfilled, we obtain the following expressions for the bulk chemical potentials of the species 
$\mu_c=k_{B}T\ln\left({\phi_0}/(1-2\phi_0)\right)$, $\mu_p=-k_{B}T\ln\left(1-2\phi_0\right)$,
where $\phi_0=n_0v$ is the bulk volume fraction of the monomeric units and counterions. The boundary conditions for the polymer concentration and electrostatic potential are~\cite{netz2003neutral,landau2013electrodynamics} $n_{p}|_s=0$, $\psi|_{s}=\psi_0$, where the symbol $|_{s}$ means that the variables are calculated at the surfaces of immersed macroscopic conductors. These boundary conditions mean that near the surface of a conductive wall (with a fixed surface potential, $\psi_0$) the monomeric units are exposed to a strong repulsive force~\cite{netz2003neutral}. Note that for simplicity we neglect the specific adsorption of the counterions. The latter can be easily taken into account~\cite{budkov2018theory}. 

Turning to the theory of macroscopic forces, let us subject the system to a dilation transformation, $x_{i}^{\prime}=x_{i}+u_{i}(\bold{r})$, where $u_i(\bold{r})$ are some arbitrary functions of coordinates. Using eqs. (\ref{EL_eq}) and assuming that the dilation is rather small, we obtain (for technical details, see Appendix)
\begin{equation}
\delta\Omega =\int d\bold{r} u_{ik}\sigma_{ik},
\end{equation}
where $u_{ik}=(\partial_{i}u_k + \partial_k u_{i})/2$ is the strain tensor~\cite{landau1986theory} and
\begin{equation}
\sigma_{ik}=\frac{\delta \Omega}{\delta u_{ik}}=\omega\delta_{ik}-\partial_{i}n_{p}^{1/2}\frac{\partial \omega}{\partial(\partial_{k}n_{p}^{1/2})}-\partial_{i}\psi\frac{\partial \omega}{\partial(\partial_{k}\psi)}
\end{equation}
is the stress tensor satisfying the local mechanical equilibrium condition, i.e.
\begin{equation}
\label{mech_eq}
\partial_{i}\sigma_{ik}=0.
\end{equation}
Using eq. (\ref{omega_2}) and excluding the bulk chemical potentials from the final expressions based on eqs. (\ref{scf_eq}), we obtain
\begin{equation}
\label{stress}
\sigma_{ik}=\sigma^{(h)}_{ik}+\sigma^{(M)}_{ik}+\sigma_{ik}^{(c)},
\end{equation}
where
\begin{equation}
\sigma_{ik}^{(h)}=-P\delta_{ik}
\end{equation}
is the standard hydrostatic stress tensor with the local pressure $P=n_p\bar{\mu}_p+n_c\bar{\mu}_c-f=-{k_{B}T}v^{-1}\left(\ln\left(1-\phi_p-\phi_c\right)+\phi_p\right),$
\begin{equation}
\sigma_{ik}^{(M)}=\varepsilon\left(\mathcal{E}_{i}\mathcal{E}_{k}-\frac{1}{2}\mathcal{E}^2\delta_{ik}\right)
\end{equation}
is the electrostatic contribution described by the standard Maxwell stress tensor~\cite{landau2013electrodynamics} in a continuous dielectric medium with the electrostatic field components $\mathcal{E}_i=-\partial_{i}\psi$, and
\begin{equation}
\label{conf_stress}
\sigma_{ik}^{(c)}=\frac{k_{B}Tb^2}{3}\left(\frac{1}{2}\nabla\cdot\left(n_p^{1/2}\nabla n_p^{1/2}\right)\delta_{ik}-\partial_{i}n_{p}^{1/2}\partial_{k}n_{p}^{1/2}\right)
\end{equation}
is the contribution to the total stress tensor originating from the conformational entropy of the polymer chains (we call it {\sl conformational} stress tensor). The expression for the conformational stress tensor (\ref{conf_stress}) is a new result of this work. We would like to note that the hydrostatic and Maxwell stress tensors for low-molecular weight electrolytes were recently obtained from the GTP in ref.~\cite{budkov2022modified}. Knowledge of the stress tensor at each point allows us to calculate the macroscopic force acting on a conductive or dielectric body immersed in a polyelectrolyte solution. It can be calculated as the surface integral over the area of an immersed body (see Appendix)
\begin{equation}
\label{force}
F_{i}= \oint\limits_{S} \sigma_{ik}n_{k} dS,
\end{equation}
where $n_k$ are the components of external normal and $dS$ is the elementary area. Note that in the absence of nonelectrostatic volume forces, due to the fact that the polyelectrolyte solution is in mechanical equilibrium, integration in eq. (\ref{force}) can be performed over any closed surface around a macroscopic body~\cite{budkov2022modified}.

\section{Polyelectrolyte solution in a slit charged nanopore}
Now let us consider the case when a polyelectrolyte solution is confined in a slit pore with identical electrified conductive walls placed at $z=0$ and $z=H$ ($H$ is the pore width). Assuming the polyelectrolyte solution in the pore is in equilibrium with the bulk liquid, we can write the self-consistent field equations as follows
\begin{empheq}[left=\empheqlbrace]{align}\nonumber
\label{scf_eq_2}
&\bar{\mu}_{p}(z)-\frac{k_{B}Tb^2}{6\phi_{p}^{1/2}(z)}(\phi_{p}^{1/2}(z))^{\prime\prime}+q\psi(z)=\mu_{p}\\
&\psi^{\prime\prime}(z)=-\frac{q}{\varepsilon v}\left(\phi_{p}(z)-\phi_{c}(z)\right),
\end{empheq}
with the boundary conditions $\psi(0)=\psi(H)=\psi_0$,~$\phi_{p}(0)=\phi_{p}(H)=0$; the volume fraction of the counterions can be analytically expressed via the electrostatic potential and volume fraction of monomeric units as follows $\phi_c=(1-\phi_{p}){e^{\frac{\mu_c+q\psi}{k_{B}T}}}/{\left(1+e^{\frac{\mu_c+q\psi}{k_{B}T}}\right)}$. The stress tensor value obtained above (\ref{stress}) allows us to determine the disjoining pressure in the pore. Indeed, the local mechanical equilibrium condition (\ref{mech_eq}) simplifies to
\begin{equation}
\frac{d\sigma_{zz}}{dz}=0,
\end{equation}
which yields
\begin{equation}
-\sigma_{zz}(z)=P_{b}+\Pi=const,
\end{equation}
where $\sigma_{zz}$ is the normal stress, $P_b$ is the bulk pressure and $\Pi$ is the disjoining pressure~\cite{derjaguin1987derjaguin}. Thus, determining $\sigma_{zz}$ at $z=H/2$, where $n_{p}^{\prime}(H/2)=\mathcal{E}(H/2)=0$, we obtain 
$\Pi=-\sigma_{zz}\left({H}/{2}\right)-P_b=-{k_{B}Tb^2}n_p^{\prime\prime}\left({H}/{2}\right)/{12}+P_m-P_b$, where we have introduced the pressure at the pore middle, $P_m=P\left({H}/{2}\right)$. The latter equation can be rewritten in a form that is more useful for applications without the second derivative of the polymer concentration. Using the first of the two eqs. (\ref{scf_eq_2}) at $z=H/2$, i.e. $-{k_BTb^2}n_p^{\prime\prime}\left({H}/{2}\right)/{12}=n_{pm}\left(\mu_{p}-\mu_{pm}-q\psi_m\right)$, we arrive at
\begin{equation}
\label{disj_press_2}
\Pi=n_{pm}\left(\mu_{p}-\mu_{pm}-q\psi_m\right)+P_m-P_b,
\end{equation}
where $\psi_m=\psi\left({H}/{2}\right)$,~$n_{pm}=n_p\left({H}/{2}\right)$,~$\mu_{pm}=\bar{\mu}_p\left({H}/{2}\right)$.
To calculate the disjoining pressure by eq. (\ref{disj_press_2}), first one has to solve self-consistent field equations (\ref{scf_eq_2}) and then calculate the respective variables at the midpoint of the pore.

\section{Numerical results and discussions}
\begin{figure}
\includegraphics[width=12 cm]{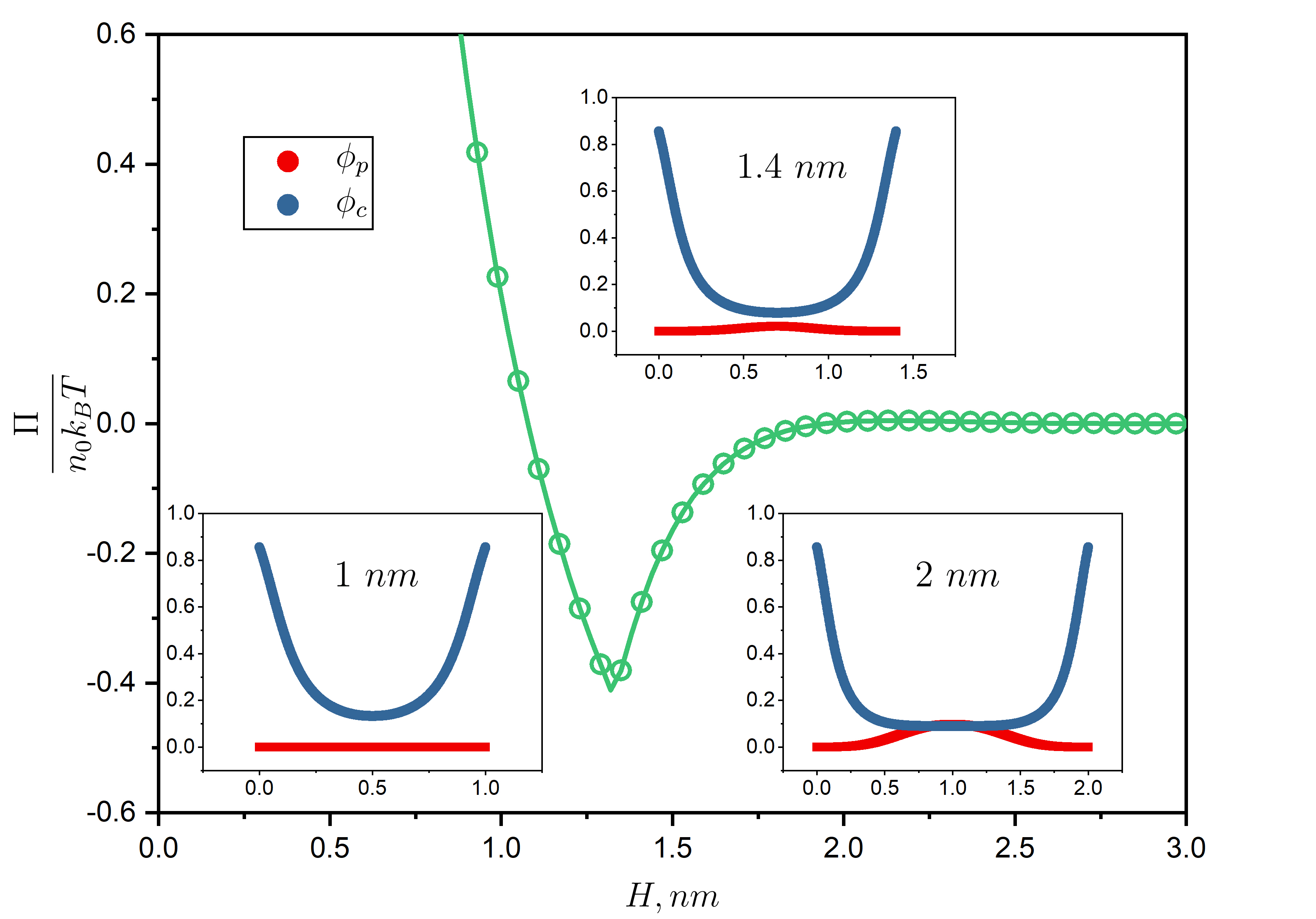}
\caption{Disjoining pressure as a function of the distance between the walls supplemented by the concentration profiles for monomeric units and counterions. The data are shown for $\phi_0=0.1$, $\psi_0=0.1~V$, $\varepsilon=40\varepsilon_0$, $b=v^{1/3}=0.5~nm$, $T=300~K$ and $q=1.6\times 10^{-19}~C$.}
 \label{fig1}
\end{figure}

\begin{figure}
 \includegraphics[width=12 cm]{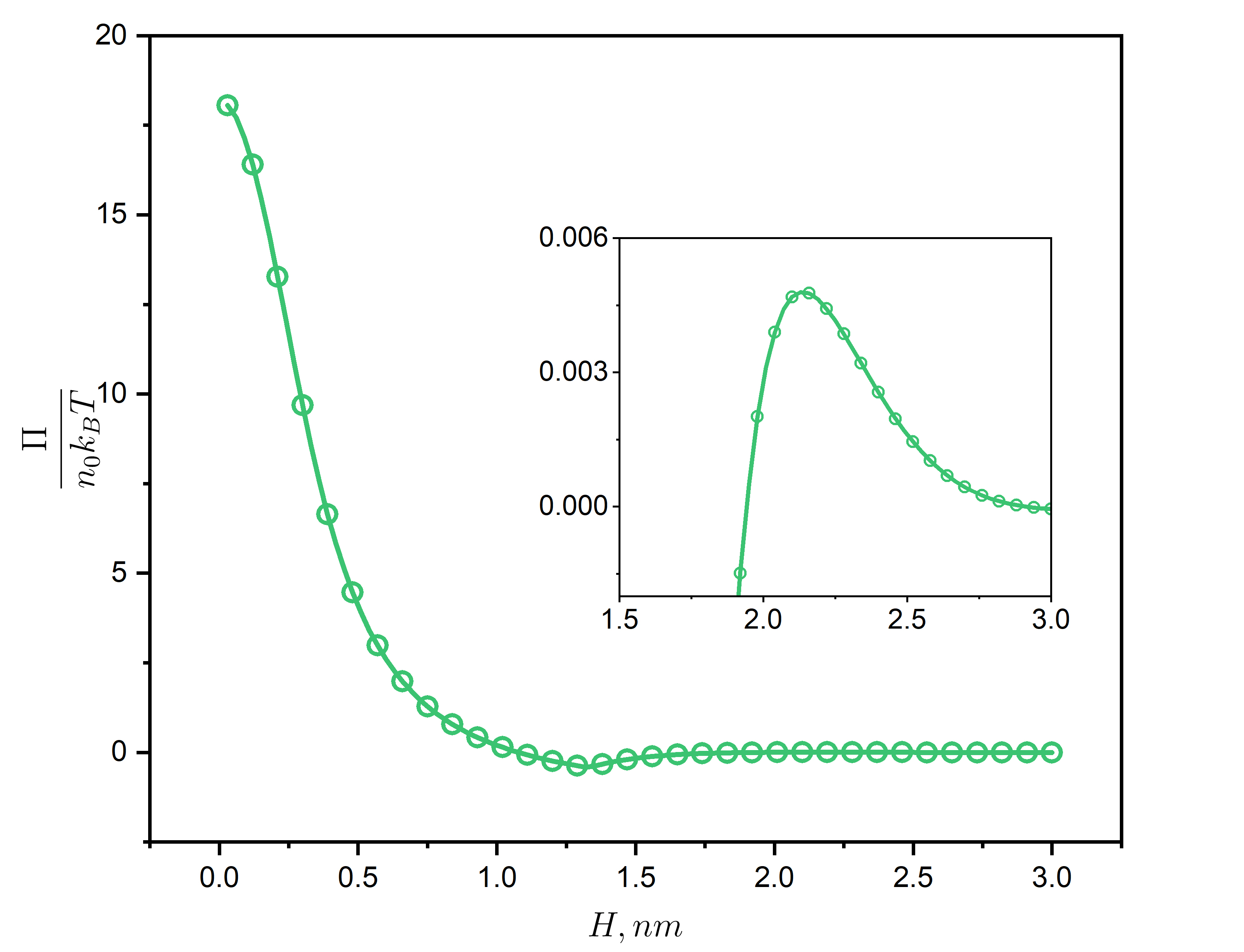}
 \caption{Disjoining pressure of PIL solution as a function of the distance between the walls. The inset shows the exponential decay at large pore widths. The data are shown for $\phi_0=0.1$, $\psi_0=0.1~V$, $\varepsilon=40\varepsilon_0$, $b=v^{1/3}=0.5~nm$, $T=300~K$ and $q=1.6\times 10^{-19}~C$.}
 \label{fig1_si}
\end{figure}

Now let us turn to the results of the numerical calculations. Fig. \ref{fig1} demonstrates the disjoining pressure of a polyelectrolyte solution in a polar solvent with the dielectric permittivity $\varepsilon=40\varepsilon_0$ (which mimics the organic solvent like dimethylsulfoxide; $\varepsilon_0$ is the empty space permittivity), fixed positive surface potential $\psi_0=0.1~V$ and bulk volume fraction $\phi_0=0.1$ as a function of the pore width; $b=v^{1/3}=0.5~nm$, $T=300~K$ and $q=1.6\times 10^{-19}~C$. The insets with the concentration profiles of the monomeric units and counterions at fixed pore widths reveal the origins of the pronounced nonmonotonic disjoining pressure behavior. In the region of sufficiently small pore widths we observe positive disjoining pressure due to the osmotic pressure of the counterions and strong overlap of the electric double layers on the charged walls. Then, at some width of the pore it turns into negative values, which is due to the smaller average counterion concentration in the middle of the pore than that in the bulk solution leading, in turn, to $P_{m}<P_b$ or $\Pi<0$ (see eq. (\ref{disj_press_2}) taking into account $n_{pm}=0$). With further increase of the pore width, the polymer chains start to permeate into the pore volume from the bulk solution (break of the disjoining pressure curve) resulting in dramatic growth in the disjoining pressure values due to a strong increase in the osmotic pressure of the counterions and monomeric units. In the case of rather wide pores the disjoining pressure behaves conventionally~\cite{derjaguin1987derjaguin}, namely, exponentially damps to zero, as in Fig. \ref{fig1_si}, where we show zoomed out disjoining pressure curve depicted at Fig. \ref{fig1}. 


\begin{figure}
 \includegraphics[width=12 cm]{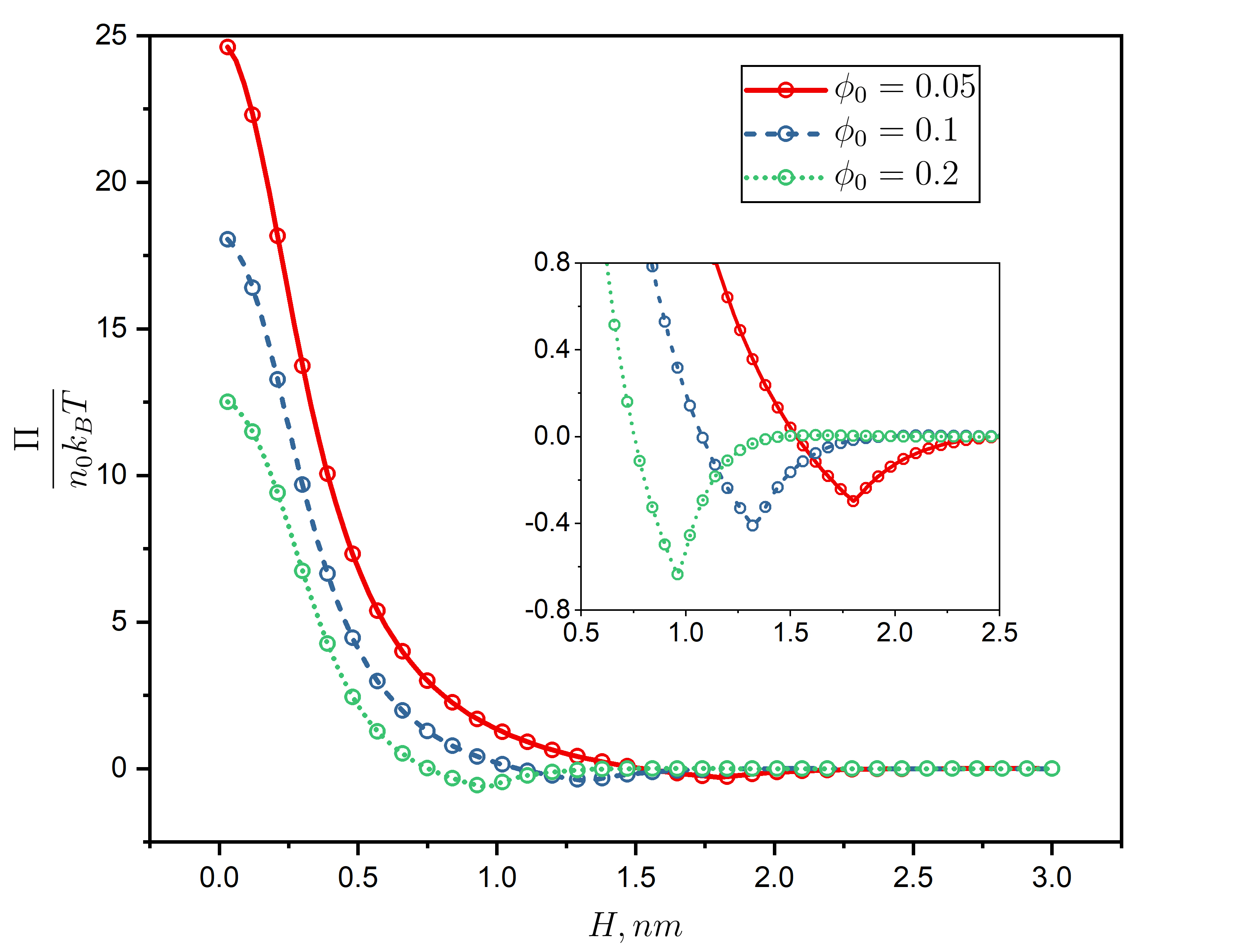}
 \caption{Disjoining pressure of polyelectrolyte solution as a function of the distance between the walls for different values of bulk volume fraction, $\phi_0$. The data are shown for $\psi_0=0.1~V$, $\varepsilon=40\varepsilon_0$, $b=v^{1/3}=0.5~nm$, $T=300~K$ and $q=1.6\times 10^{-19}~C$.}
 \label{fig2_si}
\end{figure}

\begin{figure}
 \includegraphics[width=12 cm]{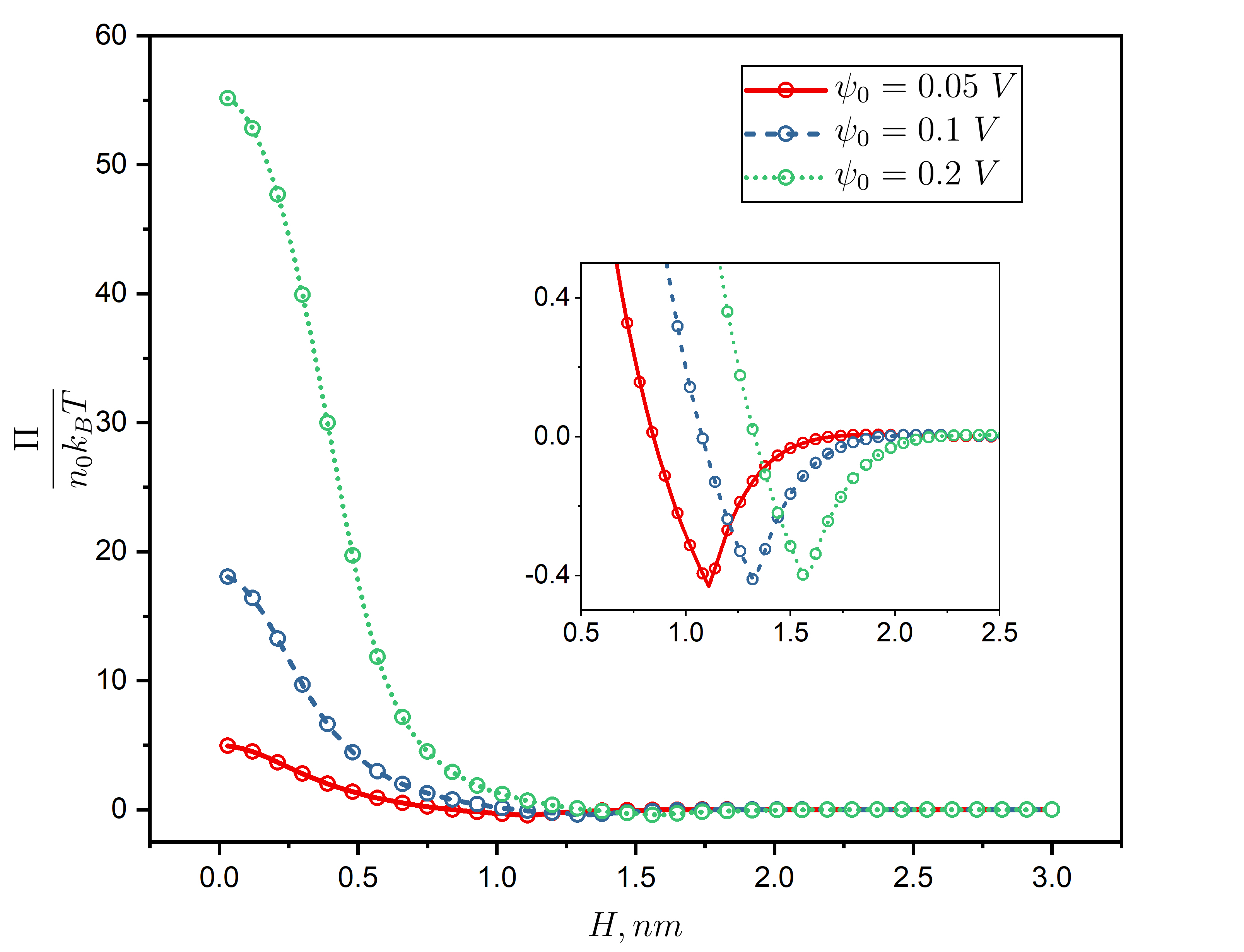}
 \caption{Disjoining pressure of polyelectrolyte solution as a function of the pore width for different values of positive surface potential $\psi_0$. The data are shown for $\phi_0=0.1$, $\varepsilon=40\varepsilon_0$, $b=v^{1/3}=0.5~nm$, $T=300~K$ and $q=1.6\times 10^{-19}~C$.}
 \label{fig3_si}
\end{figure}

\begin{figure}
\includegraphics[width=12 cm]{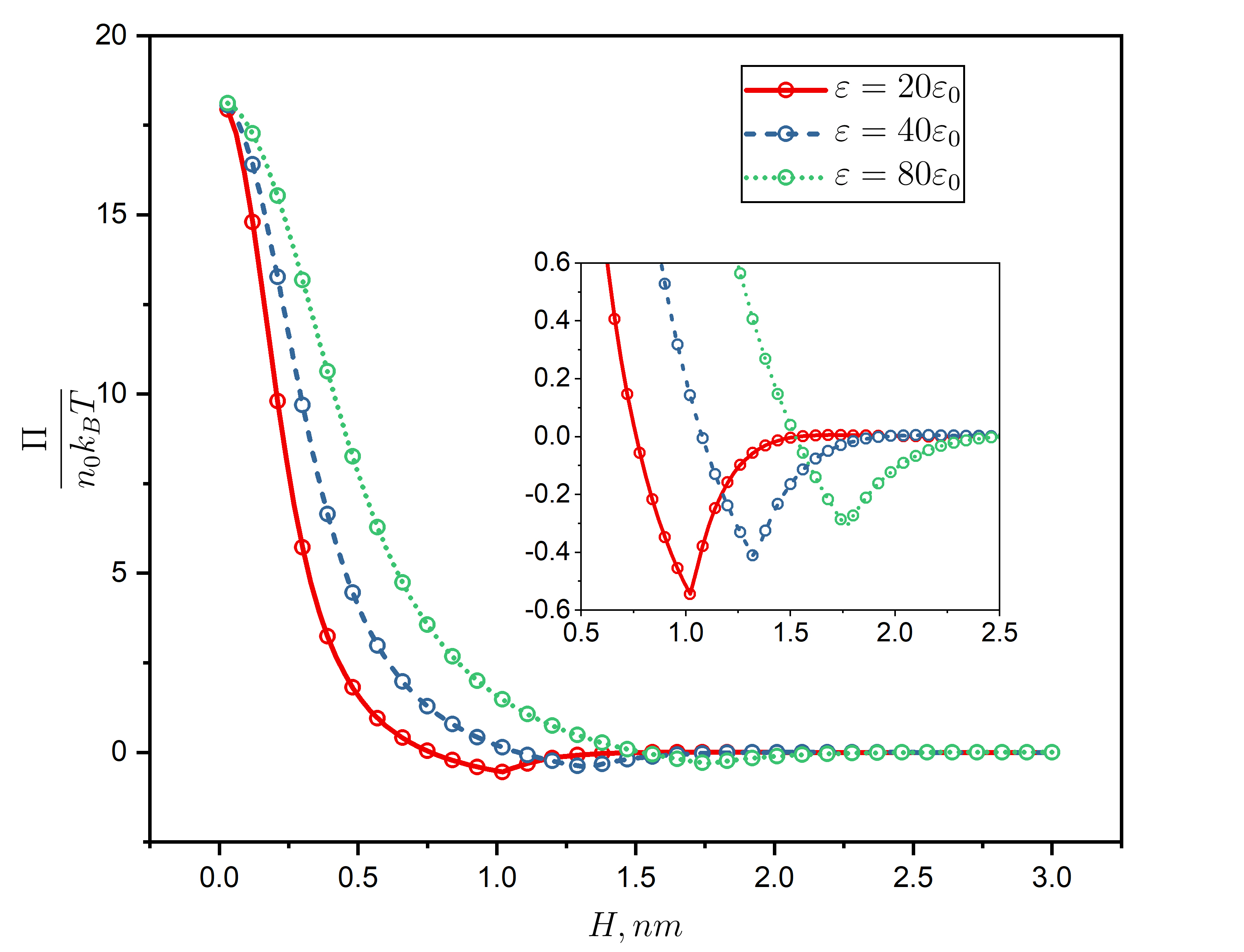}
\caption{Disjoining pressure of polyelectrolyte solution as a function of the distance between the walls for different values of dielectric constant $\varepsilon$ (in $\varepsilon_0$ units). The data are shown for $\phi_0=0.1$, $\psi_0=0.1~V$, $b=v^{1/3}=0.5~nm$, $T=300~K$ and $q=1.6\times 10^{-19}~C$.}
\label{fig4_si}
\end{figure}

The width, at which the polymer starts to permeate the pore, can be tuned by changing the bulk volume fraction, surface potential and dielectric permittivity of the solvent, which is demonstrated at Figs. \ref{fig2_si}, \ref{fig3_si} and \ref{fig4_si}, correspondingly. The insets in plots highlight the region of nonmonotonic disjoining pressure behavior. As is seen, due to the steric effect, the more the volume fraction of the monomeric units is in the bulk (Fig. \ref{fig2_si}) the faster, i.e. at less width, they enter the space between the walls, despite the same sign of the surface and monomeric unit charge. Besides, for sufficiently small pores higher bulk volume fraction leads to stronger screening of the surface potential resulting in lower disjoining pressure values and deeper minimum. With increase of positive surface potential we obtain the opposite picture (Fig. \ref{fig3_si}) -- the minimum shifts to the region of wider pores. This is simply due to enforcement of the polymer-wall electrostatic repulsion, thus moving the width, at which the first polymer pore penetration occurs to larger pores. With increase of the positive surface potential we obtain the opposite picture (Fig. \ref{fig3_si}) -- the minimum shifts to the region of wider pores. This is simply due to enforcement of the polymer-wall electrostatic repulsion, thus we need more counterions in the pore to screen it, which leads to the shift of the effect to larger pores. With increase of the solvent dielectric permittivity (Fig. \ref{fig4_si}) the screening of the surface potential decreases. It leads to growth of the counterions concentration in the middle of the slit, which in turn results in stronger osmotically-enforced interwall repulsion. Increase in the counterion concentration makes it difficult for monomeric units to penetrate the pore volume that manifest itself via the shift of the location of the minimum on the disjoining pressure profiles to larger slit widths.

\begin{figure}
\includegraphics[width=12 cm]{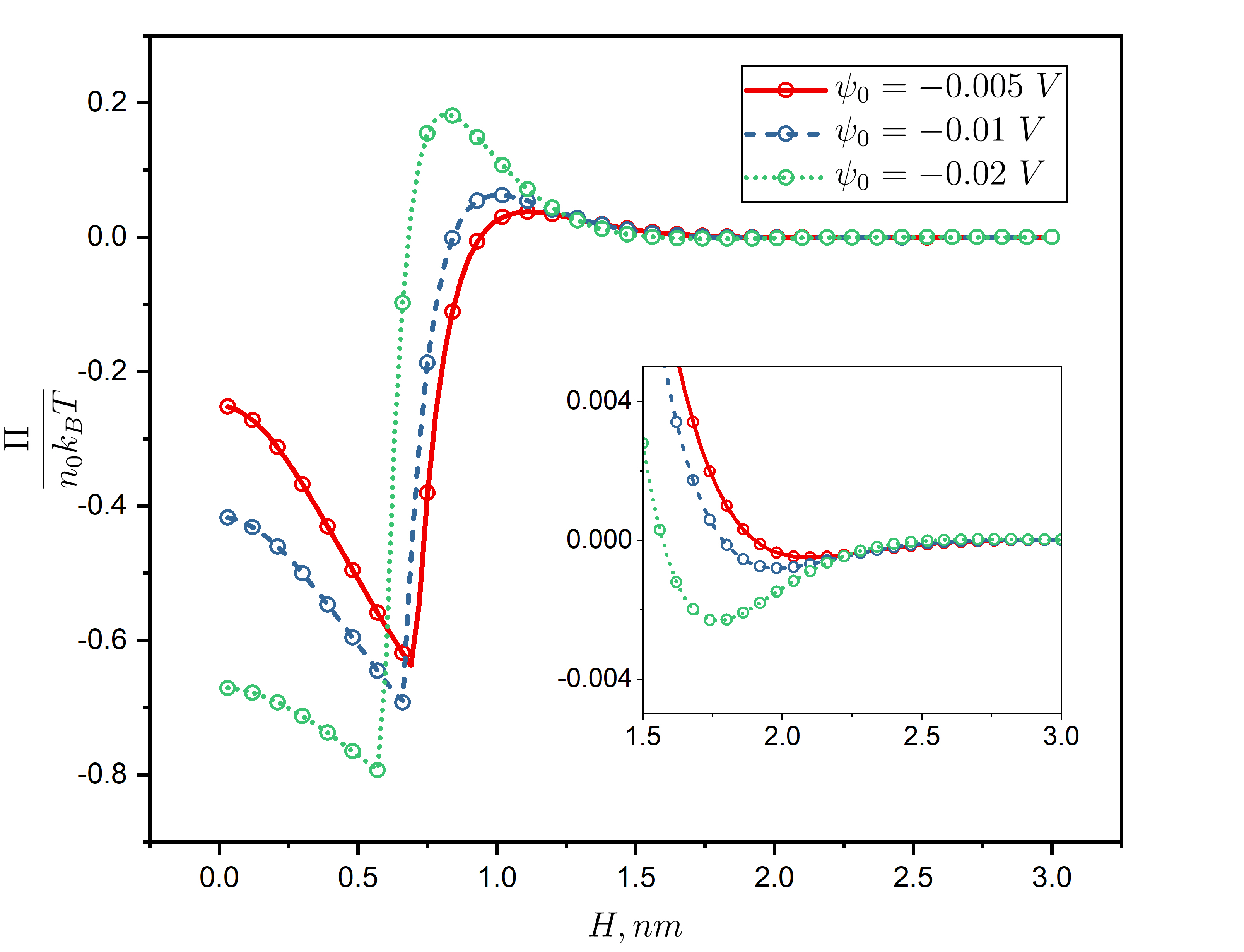}
\caption{Disjoining pressure of polyelectrolyte solution as a function of distance between the walls for different values of negatively charged surface potential $\psi_0$. The data are shown for $\phi_0=0.1$, $\varepsilon=40\varepsilon_0$, $b=v^{1/3}=0.5~nm$, $T=300~K$ and $q=1.6\times 10^{-19}~C$.}
\label{fig5_si}
\end{figure}

Fig. \ref{fig5_si} demonstrates the behavior of disjoining pressure as a function of slit separation for different values of negative surface potentials. At small pore widths concentration of counterions in the pore is significantly lower than that in the bulk due to surface-counterions electrostatic repulsion, which leads to negative disjoining pressure values. Similar disjoining pressure behavior was obtained in paper~\cite{landman2021repulsive} for salt polyelectrolyte solution confined in a slit pore between two charged dielectric membranes.Starting fFrom the width, where the monomeric units of polyelectrolytes start to permeate the pore volume we observe a drastic growth of disjoining pressure, revealing oscillation behavior as in paper~\cite{podgornik2004polyelectrolyte} where the author considered the bridging interaction of the colloid particles, provided by the oppositely charged polyelectrolyte chain, adsorbed on them.

\begin{figure}
 \includegraphics[width=12 cm]{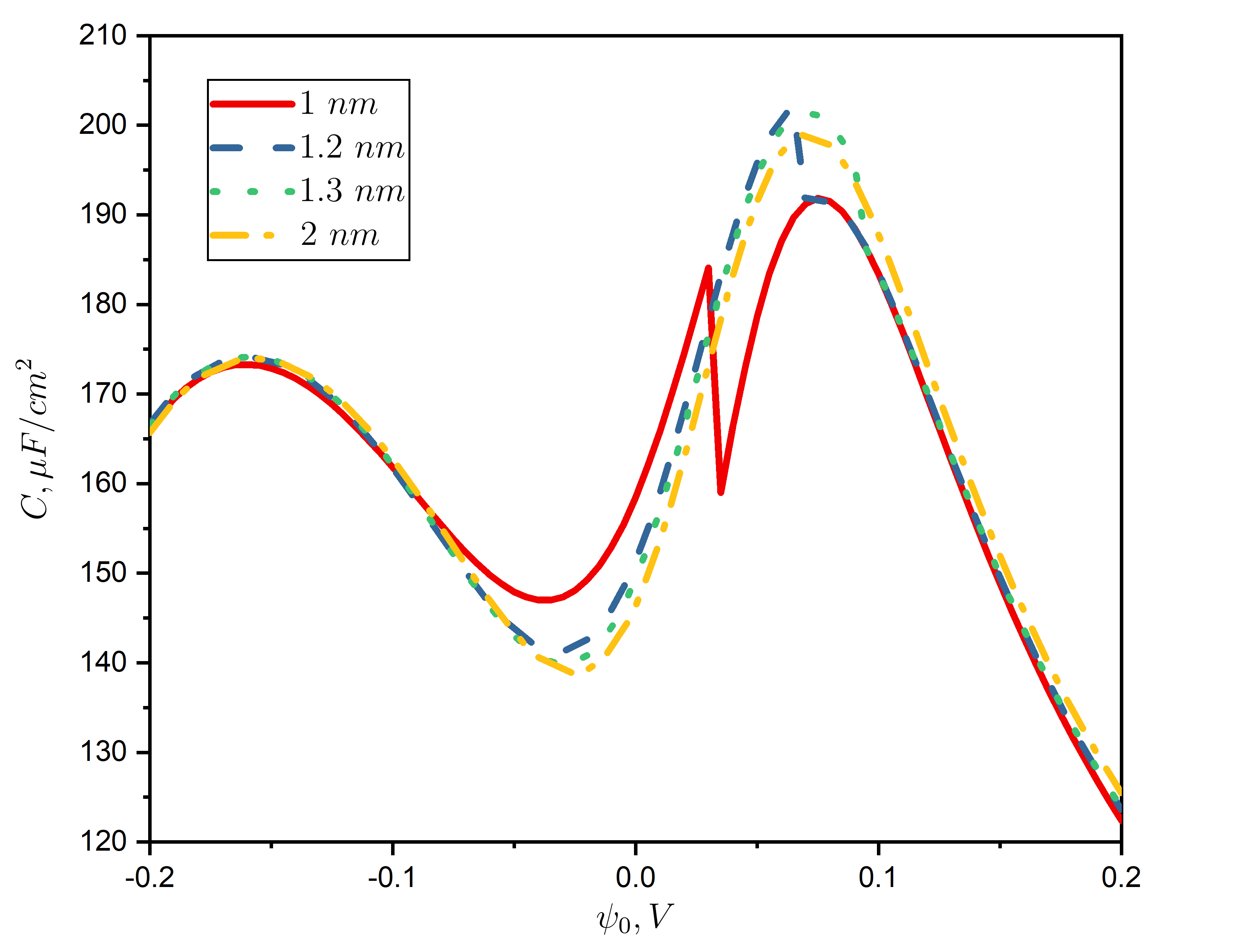}
 \caption{Differential capacitance profiles plotted for different pore widths. The data are shown for $H=1~nm$, $\phi_0=0.1$, $\varepsilon=40\varepsilon_0$, $b=v^{1/3}=0.5~nm$, $T=300~K$ and $q=1.6\times 10^{-19}~C$.}
 \label{fig2}
\end{figure}

\begin{figure}
 \includegraphics[width=12 cm]{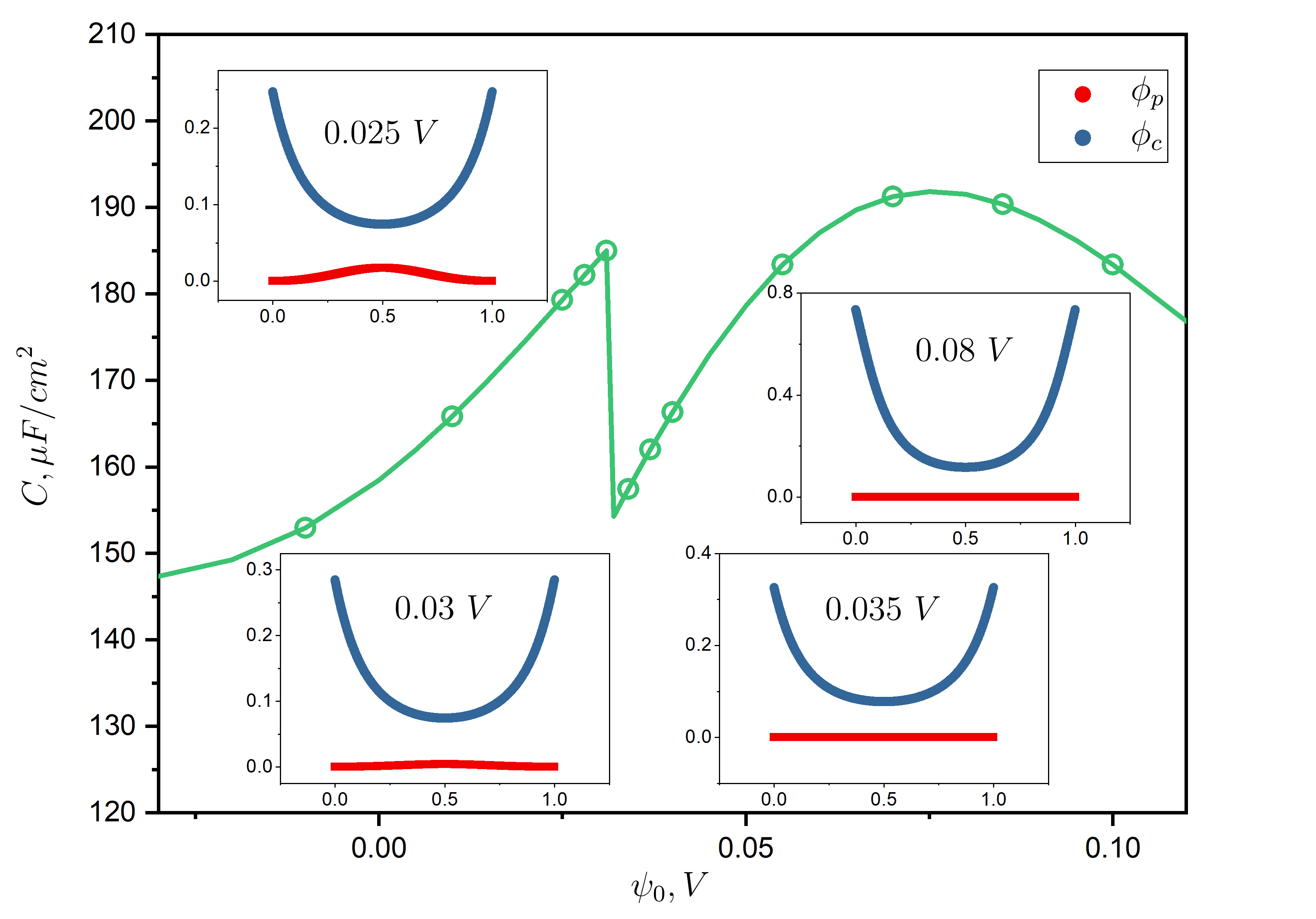}
 \caption{Differential capacitance profile for a $1~nm$ pore supplemented by the concentration profiles at certain surface potentials. The data are shown for $\phi_0=0.1$, $\varepsilon=40\varepsilon_0$, $b=v^{1/3}=0.5~nm$, $T=300~K$ and $q=1.6\times 10^{-19}~C$.}
 \label{fig3}
\end{figure}

The effect discovered for the disjoining pressure might as well be observed for differential capacitance, $C=d\sigma/d\psi_0$, as a function of the surface potential ($\sigma=-\varepsilon \psi^{\prime}(0)$ is the surface charge density of the pore walls)~\cite{kornyshev2007double,budkov2021electric}. For small pores (Fig. \ref{fig2}, the parameters are the same as above), the differential capacitance profile demonstrates similar abrupt nonmonotonic behavior, then, as the pores become wider, we observe a satellite peak, which then flattens out resulting in a profile similar to the one obtained for an isolated electric double layer at the interface of a polyelectrolyte solution/charged electrode~\cite{budkov2022electrochemistry}. The nature of such pronounced nonmonotonic behavior of the differential capacitance curve for a $1~nm$ pore is illustrated in the same manner as above (see the insets in Fig. \ref{fig3}). The abrupt drop in the differential capacitance is associated with the electrostatic repulsion-driven exclusion of the polymers from the pore at a certain positive surface potential value; the following capacitance growth is determined by the increase in the counterion concentration in the pore with the increase in the surface potential.

\begin{figure}
 \includegraphics[width=12 cm]{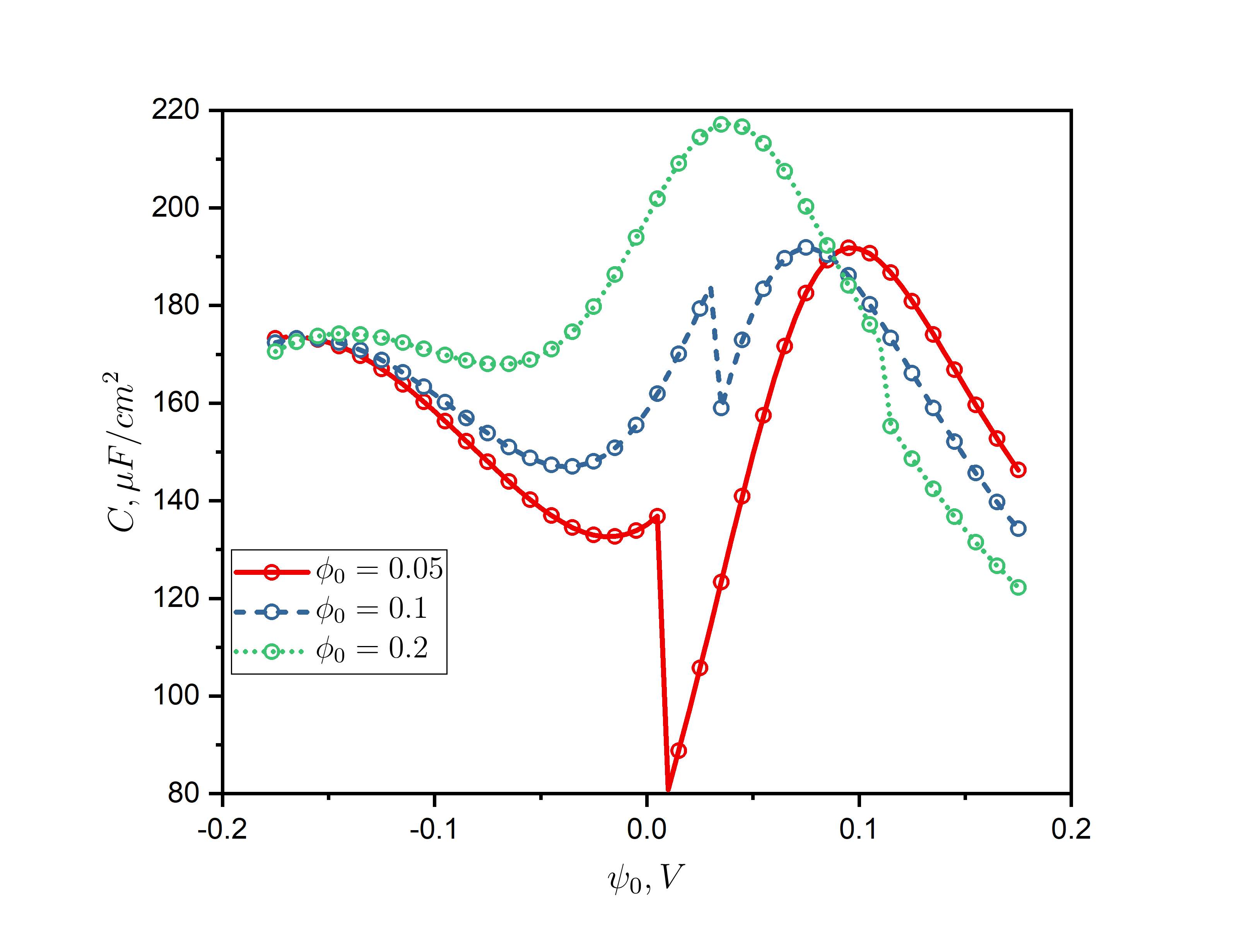}
 \caption{Differential capacitance profiles for different bulk volume fractions. The data are shown for $H=1~nm$, $\varepsilon=40\varepsilon_0$, $b=v^{1/3}=0.5~nm$, $T=300~K$ and $q=1.6\times 10^{-19}~C$.}
 \label{fig4}
\end{figure}

\begin{figure}
 \includegraphics[width=12 cm]{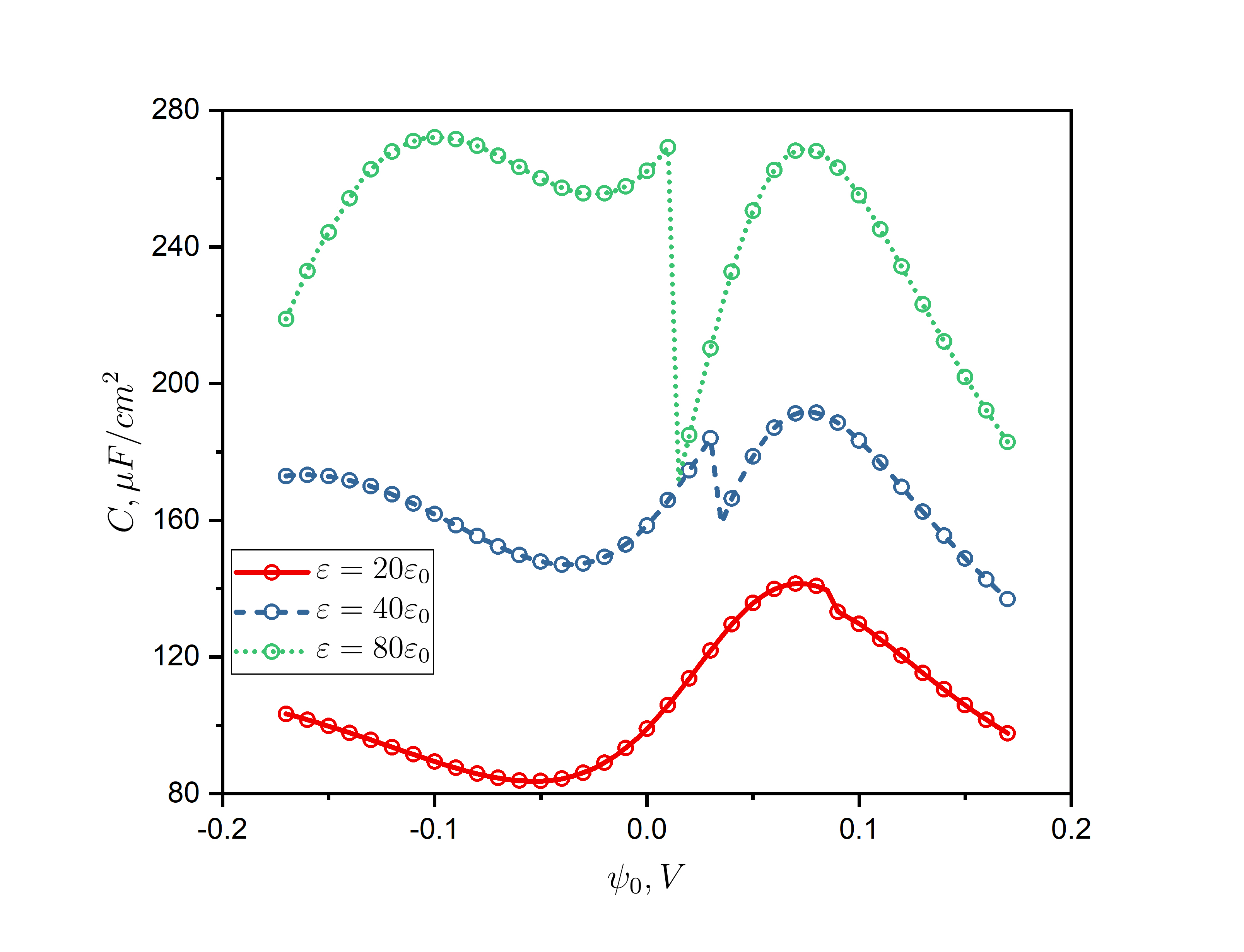}
 \caption{Differential capacitance profiles for different dielectric constants $\varepsilon$ (in $\varepsilon_0$ units). The data are shown for $H=1~nm$, $\phi_0=0.1$, $b=v^{1/3}=0.5~nm$, $T=300~K$ and $q=1.6\times 10^{-19}~C$.}
 \label{fig5}
\end{figure}

Moreover, similar to the disjoining pressure case, the nonmonotonic behavior of the differential capacitance profile can be altered by the change of the system parameters. In such a way, Fig. \ref{fig4} demonstrates how the change of the bulk volume fraction influences the shape of the differential capacitance profile for the pore of $1~nm$ width. At negative surface potentials bulk volume fraction increase simply means that more species can permeate the pore volume leading to higher charge accumulation and the growth of the capacitance values. With the change of sign of the surface potential values, we start to observe an abrupt drops of the differential capacitance profiles due to the fact that the counterions screening of the surface potential becomes insufficient to keep the polymer in the pore volume due to the electrostatic repulsion. Thus, the clear tendency is as follows: the less the bulk volume fraction, the less counterions in the pore screen the surface potential, the lower positive value of the surface potential is needed to eject the polymer from the pore volume. Differential capacitance profiles also demonstrate sensitivity to the solvent dielectric permittivity change (see Fig. \ref{fig5}). The interpretation is the same as for the disjoining pressure -- with increase in dielectric permittivity the screening of the surface potential decreases. Thus, for the negative surface potentials electrostatic screening reduction leads to the drastic increase in the polymer concentration in the pore, and in turn results in considerable  differential capacitance growth. In the region of positive surface potential, the more polar the solvent the lower potential value, which provides the polymer expelling from the pore volume.

\section{Concluding remarks}
To sum it up, we formulated a self-consistent field theory of macroscopic forces in spatially inhomogeneous equilibrium polyelectrolyte solutions, deriving the total stress tensor consistent with self-consistent field equations. Alongside the expected contributions obtained earlier for low-molecular weight electrolytes~\cite{budkov2022modified} (hydrostatic and Maxwell stress tensors), the total stress tensor for polyelectrolyte solutions contains an additional term linked to the conformational entropy of  flexible polymer chains -- conformational stress. Based on the obtained total stress tensor, we calculated the disjoining pressure between two identical conductive walls immersed in a polyelectrolyte solution. We discovered singular behavior of the disjoining pressure as a function of slit separation for the case when the signs of the macroions and wall charges are the same and investigated its nature by analyzing the concentration profiles of the macroions and counterions. We also showed the influence of the discovered effect on differential capacitance demonstrating its jump-like behavior. The observed effect can be tuned to appear at physically reasonable parameters, thus, it can be potentially registered experimentally. In practice it can be achieved for materials with significantly narrow unimodal pore size distribution, otherwise the effect can be smoothed out. 

In conclusion, we would like to rise some issues regarding further development of the proposed approach. Firstly, it is remained unanswered, how this theory can be implemented to the polymeric systems with more complicated chain structures, like copolymer composition, chain architectures and charge distribution. For this purpose, it is necessary to know the functional of the conformational free energy, similar to Lifshitz one for the flexible Gaussian chains, for each polymeric system. Secondly, it is important to incorporate presented theoretical approach into more sophisticated polymeric theories beyond the ground state dominance approximation, such as the self-consistent field theory~\cite{fredrickson2006equilibrium} and the Lifshitz theory~\cite{khokhlov1994statistical}.

\acknowledgements
The authors thank A.L. Kolesnikov for fruitful discussions and help with numerical calculations. The authors thank the Russian Science Foundation (Grant No. 22-13-00257). This research was supported in part by the computational resources of the HPC facilities at HSE University~\cite{hse_cluster}.

\section*{Appendix}

The grand thermodynamic potential (GTP) of the polyelectrolyte solution is 
\begin{equation}
\Omega=\int d\bold{r}\omega(\bold{r}),  
\end{equation}
where the GTP density is 
\begin{equation}
\omega(\bold{r})=\omega(\psi(\bold{r}),\nabla\psi(\bold{r}),\xi(\bold{r}),\nabla\xi(\bold{r}),n_c(\bold{r}),\bold{r})
\end{equation}
with $\xi(\bold{r})=n_{p}^{1/2}(\bold{r})$.
Let us subject the system to a dilation transformation, $x_{i}^{\prime}=x_{i}+u_{i}(\bold{r})$. The variation of the GTPs is
\begin{equation}
\delta\Omega=\int d\bold{r}^{\prime} \omega^{\prime}(\bold{r}^{\prime})-\int d\bold{r}\omega(\bold{r}),
\end{equation}
where 
\begin{equation}
\omega^{\prime}(\bold{r}^{\prime})=\omega(\psi^{\prime}(\bold{r}^{\prime}),\nabla^{\prime}\psi^{\prime}(\bold{r}^{\prime}),\xi^{\prime}(\bold{r}^{\prime}),\nabla^{\prime}\xi^{\prime}(\bold{r}^{\prime}),n_c^{\prime}(\bold{r}^{\prime}),\bold{r}^{\prime})
\end{equation}
Assuming that the absolute value of the vector field $\bold{u}$ is sufficiently low, we obtain a linear approximation in $u_i$
\begin{equation}
\delta\Omega = \int d\bold{r}\left(1+u_{ii}\right)\left(\omega^{\prime}(\bold{r})+u_{i}\partial_{i}\omega^{\prime}(\bold{r})\right)-\int d\bold{r}\omega(\bold{r})=\int d\bold{r}\left(\omega^{\prime}(\bold{r})-\omega(\bold{r})+\partial_{i}\left(u_i\omega(\bold{r})\right)\right),   
\end{equation}
where $u_{ii}=\partial_{i}u_i$ is the deformation tensor trace ~\cite{landau1986theory}, $u_{ij}=\left(\partial_{i}u_{j}+\partial_{j}u_{i}\right)/2$; we have taken into account that in the linear approximation $u_i\omega^{\prime}(\bold{r})\approx u_i\omega(\bold{r})$. Then, we arrive at
\begin{equation}
\omega^{\prime}(\bold{r})-\omega(\bold{r})=\frac{\partial \omega}{\partial\psi}\delta\psi+\frac{\partial \omega}{\partial(\partial_{i}\psi)}\partial_i\delta\psi+\frac{\partial \omega}{\partial\xi}\delta\xi+\frac{\partial \omega}{\partial(\partial_{i}\xi)}\partial_i\delta\xi=\partial_{i}\left(\frac{\partial \omega}{\partial(\partial_{i}\psi)}\delta\psi+\frac{\partial \omega}{\partial(\partial_{i}\xi)}\delta\xi\right),
\end{equation}
where we use the Euler-Lagrange equations
\begin{equation}
\label{EL_eq_2}
\frac{\partial{\omega}}{\partial{\psi}}=\partial_{i}\frac{\partial \omega}{\partial(\partial_{i}\psi)},~\frac{\partial{\omega}}{\partial{\xi}}=\partial_{i}\frac{\partial \omega}{\partial(\partial_{i}\xi)},~\frac{\partial{\omega}}{\partial{n_{c}}}=0.
\end{equation}
As already noticed in the main text, we adopted the Einstein rule implying the summation over the repeated indices. Taking into account that $\delta\psi(\bold{r})=\psi^{\prime}(\bold{r})-\psi(\bold{r})=-u_k\partial_k \psi(\bold{r})$ and $\delta\xi(\bold{r})=\xi^{\prime}(\bold{r})-\xi(\bold{r})=-u_k\partial_k \xi(\bold{r})$, we arrive at
\begin{equation}
\label{variation}
\delta\Omega = \int d\bold{r}\partial_i\left(u_k\sigma_{ik}\right),    
\end{equation}
where 
\begin{equation}
\sigma_{ik}=\omega\delta_{ik}-\frac{\partial \omega}{\partial(\partial_{i}\psi)}\partial_k\psi-\frac{\partial \omega}{\partial(\partial_{i}\xi)}\partial_k\xi
\end{equation}
is some symmetric tensor. In the case, when $\omega$ does not explicitly depend on coordinates, i.e. $\omega(\bold{r})=\omega(\psi(\bold{r}),\nabla\psi(\bold{r}),\xi(\bold{r}),\nabla\xi(\bold{r}),n_c(\bold{r}))$, as it follows from the Euler-Lagrange equations (\ref{EL_eq_2}) the tensor $\sigma_{ik}$ is divergenceless, i.e. $\partial_i\sigma_{ik}=0$. Therefore, we obtain the following expression 
\begin{equation}
\delta\Omega =\int d\bold{r} u_{ik}\sigma_{ik}
\end{equation}
from which we conclude that $\sigma_{ik}$ is nothing more than the stress tensor, i.e.
\begin{equation}
\sigma_{ik}(\bold{r})=\frac{\delta\Omega}{\delta u_{ik}(\bold{r})}.
\end{equation}
Using the divergence theorem we obtain
\begin{equation}
\label{variation}
\delta\Omega=\oint\limits_S dSu_kn_{i}\sigma_{ik}=\oint\limits_S dS P_ku_k,    
\end{equation}
where integration is performed over surface of an immersed in solution body, $n_k$ are the components of external normal and $dS$ is the elementary area. As it follows from eq. (\ref{variation}), variation of the GTP is mechanical work of the body deformation under external force with surface density $P_{k}=n_{i}\sigma_{ik}$. Thus, knowledge of the stress tensor at each point allows us to calculate the macroscopic force acting on a conductive or dielectric body immersed in a polyelectrolyte solution
\begin{equation}
\label{force2}
F_{i}= \oint\limits_{S} P_i dS.
\end{equation}

\bibliographystyle{vancouver}
\bibliography{name}
\end{document}